# Extraordinary wavelength reduction in terahertz graphene-cladded photonic crystal slabs

Ian A. D. Williamson[1], S. Hossein Mousavi[1], Zheng Wang[1,*]

[1]Microelectronics Research Center, Department of Electrical and Computer Engineering, The University of Texas at Austin, Austin, TX 78758 US

[*]Corresponding author: zheng.wang@austin.utexas.edu

## Abstract

Photonic crystal slabs have been widely used in nanophotonics for light confinement, dispersion engineering, nonlinearity enhancement, and other unusual effects arising from their structural periodicity. Sub-micron device sizes and mode volumes are routine for silicon-based photonic crystal slabs, however spectrally they are limited to operate in the near infrared. Here, we show that two single-layer graphene sheets allow silicon photonic crystal slabs with submicron periodicity to operate in the terahertz regime, with an extreme 100x wavelength reduction and excellent out-of-plane confinement. The graphene-cladded photonic crystal slabs exhibit band structures closely resembling those of ideal two-dimensional photonic crystals, with broad two-dimensional photonic band gaps even when the slab thickness approaches zero. The overall photonic band structure not only scales with the graphene Fermi level, but more importantly scales to lower frequencies with reduced slab thickness. Just like ideal 2D photonic crystals, graphene-cladded photonic crystal slabs confine light along line defects, forming waveguides with the propagation lengths on the order of tens of lattice constants. The proposed structure opens up the possibility to dramatically reduce the size of terahertz photonic systems by orders of magnitude.

## Introduction

During the past two decades, photonic crystals[1,2] have revolutionized the field of nanophotonics, as their periodically modulated structure gives rise to a wide range of unprecedented phenomena in manipulating the flow of light[3,4]. Photonic crystal slabs[5–7], in particular, have emerged as one of the most practical variety of photonic crystals because their 2D periodicity and planar structure is highly compatible with standard lithographic fabrication processes[8,9]. Photonic crystal slabs have periodicity only two of their in-plane directions, and rely on total internal reflection, i.e. index guiding, to confine light in the third out-of-plane dimension. Although photonic crystals with fully-3D periodicity are capable of providing a complete 3D photonic band gap and additional degrees of freedom for tuning band dispersion, their fabrication is considerably more challenging over photonic crystal slabs, especially for obtaining point and line defects at optical frequencies[10]. Consequently, photonic crystal slabs have been deployed in a wide range of applications, most prominently as waveguides and resonators with very strong field confinement and high quality factors[11,12]. They have also enabled large spatial dispersion for negative refraction and self-collimation[13–15] and are capable of supporting large temporal dispersion for engineered slow light waveguides[16,17]. Additionally, they have recently enabled strong photon-phonon coupling[18] and 2D topological orders for disorder-immune photonic edge states[19–23].

Unfortunately, a fundamental limitation of photonic crystal slabs is that their periodicity is proportional to the operating free-space wavelength. This translates to device sizes and mode volumes that are only slightly below the diffraction limit, even when materials with a high refractive index are used. In a silicon photonic crystal slab for instance, the lattice constant defining the in-plane periodicity is generally only about 0.30 to 0.40 times the free space wavelength[3,5]. For example, although high-resistivity float-zone silicon is transparent around 1 THz, *submicron periodicity* in silicon photonic crystal slabs would be too small to provide a band gap. Additionally, because total internal reflection provides the out-of-plane confinement, only an incomplete band gap is obtained due to the presence of radiation modes that extend into the substrate and superstrate. The size of the in-plane band gap is also significantly reduced from that of a purely 2D crystal, because of the incomplete overlap between the Bloch mode and the slab. Additionally, out-of-plane radiation losses limit the quality factor of point-defect resonators[12]. Thus, an alternative means to further the wavelength reduction and out-of-plane light confinement is highly desirable for miniaturization and system integration.

One well-known solution for better confinement is to employ a 3D structure that is uniform in the out-of-plane direction and sandwiched between two truncating reflecting boundaries, which are closely spaced such that only the lowest-order mode with a pure polarization is supported (e.g. the TM polarization with electric fields out-of-plane when perfect electrical conductors are used). Widely exploited in microwave photonic crystals, this approach has been realized by sandwiching a 2D slab between two metal plates that form a parallel plate waveguide[24,25]. Parallel plate waveguides support a fundamental TEM mode with linear dispersion, no low frequency cutoff, and no radiation losses. The combination of the strongly confined TEM mode and the modulated intermediate dielectric produces band structures that are identical to the TM modes in purely 2D photonic crystals. The microwave regime is uniquely suited for this approach because highly conductive metals are readily available and the macroscopic feature sizes of these low-frequency crystals are easy to fabricate. However, at optical frequencies, it becomes challenging to realize ideal parallel-plate waveguides due to the fact that noble metals, such as copper and gold, are plasmonic with significantly increased ohmic loss. Additionally, with significant field penetration into the plasmonic material, less of the guided mode is able to interact with the modulated dielectric in optical parallel plate waveguides.

Recently, graphene has emerged as an excellent conductor for the parallel plate geometry at THz frequencies: a quasi-TEM mode is largely confined between two graphene sheets[26] and a wavelength that is over 40 times smaller than in free-space[27], thanks to a large kinetic inductance.

Although occasionally referred as the transverse magnetic graphene plasmon[28], the quasi-TEM mode is qualitatively different from the graphene plasmon, and is free from the associated drawbacks: a nonlinear $k \sim \omega^2$ dispersion, a radiative low frequency cutoff, and a very non-uniform field distribution where the mode is tightly confined to the graphene surface.[29–32] Moreover, the graphene parallel plate waveguide TEM mode has an absorption-limited decay length that is capable of exceeding tens of wavelengths, and does not experience a skin effect that is typically seen in conventional metals[33]. If the mode energy resides predominantly in the space between the graphene sheets, it stands to reason that the graphene TEM mode with a modulated dielectric structure represents an ideal platform to realize of a purely 2D TM system with excellent out-of-plane confinement, and the added benefit of radical reduction in wavelength.

In this paper, we combine the terahertz graphene parallel plate waveguide and the photonic crystal slab to realize extremely-subwavelength guided Bloch modes with strong out-of-plane confinement. Here two parallel graphene sheets are necessary, and they behave very differently from photonic crystal slabs interfacing with only one graphene sheet[34], or 1D photonic crystals where graphene is used for absorption[35] and magneto-optical effects[36]. The graphene parallel plates are continuous and have no spatial modulation, unlike systems with isolated graphene islands arranged in 2D patterns to modify the photonic band structures[37,38] or to enhance absorption[39–41]. Also, the extreme wavelength reduction and the TM-like mode profile are absent from previous work in integrating graphene sheets with photonic crystal slabs at infrared or visible frequencies, to realize tunability[42,43] and nonlinearity[44], to dissipate heat[45], and to acoustically steer an beam[46].

We begin this article by introducing the physical configuration and discussing the unique band structure of the graphene photonic crystal slab. We also examine the factors influencing the attenuation of the guided modes. We then discuss the effect that the structure's thickness and graphene Fermi level have on the size and location of the photonic band gap. Next, we examine the conditions under which the structure exhibits scale invariance that is consistent with all-dielectric photonic crystals. We then discuss the structure's ability to support modes which are confined to defect waveguides in the bulk crystal. Finally, we conclude by discussing the device's prospects as a building block for more complex applications.

## Results

### Subwavelength band structures of the graphene-cladded photonic crystal slab

For concreteness, the structure which we consider is a triangular lattice of silicon rods $\left(\text{relative } \varepsilon = 11.67\right)$[47] embedded in a low index material $\left(\text{relative } \varepsilon = 2.25\right)$, such as glass or polymer[48,49]. An exploded schematic of the structure is shown in Fig. 1(a) where the in-plane periodicity (*a*) is 300 nm and the slab thickness (*g*), or equivalently, graphene separation, is 40 nm. The two single layer graphene sheets have identical properties and are characterized by a modified Drude surface conductivity given by[29,50]

$$\sigma(\omega) = -j \frac{e^2 k_b T}{\pi \hbar^2} \frac{1}{(\omega - j\gamma)} \left[ \frac{E_F}{k_b T} + 2\log\left(e^{-E_F/k_b T} + 1\right) \right], \qquad (1)$$

which is precise in the THz/sub-THz spectral range. In this work we focus on highly doped graphene, with a Fermi level exceeding 0.1 eV, and a scattering rate of 0.4 meV, which is within reach of state of the art fabrication techniques[32]. With such high doping levels, our chosen operating frequency range is far below the onset of interband transitions, and thus the effect of room temperature, i.e. washing out of the interband transition edge, is negligible.

First-principle numerical modeling confirms that the mode profile of the graphene-cladded photonic crystal slab is very close to that of a 2D crystal[3]. Fig. 1(b) illustrates the magnitude of the electric field across two orthogonal planes which cut through the symmetry planes of the 3D unit cell of our structure. We first observe in the vertical cut (Fig. 1(b), left) that the majority of the mode is concentrated within the slab and between the graphene sheets. It is also apparent that the magnitude of the field between the sheets is nearly uniform in the vertical direction which agrees with the predicted behavior of the TEM mode (which also has negligible in-plane electric field components). Furthermore, we see in the horizontal cut midway through the slab (Fig. 1(b), right), that the mode pattern resembles the unit-cell field pattern for the lowest band of a purely 2D TM crystal, in which an increased field concentration is encountered in the high-index silicon rod.

Remarkably, the calculated band structure in Fig. 1(c) shows that the addition of the graphene opens up a wide band gap in a slab that would be otherwise too thin to support any gap. In fact, the relative size of the gap for the TEM mode is nearly equal to that of the 2D TM band gap (shown in Fig. S1), which is approximately 28%. In principal, the graphene-cladded photonic crystal slab can replicate any feature of the purely 2D band structure, including a second photonic band gap (which occurs at higher frequencies). However, a larger refractive index contrast between the rod and cladding materials is be needed to support a second band gap. Simply replacing the oxide cladding with air to leave free-standing silicon rods may increase fabrication complexity, thus a square lattice of air holes in silicon may be a better approach for obtaining a second band gap.

A notable feature of the graphene-cladded photonic crystal slab is that the lattice constant and Bloch modes have a spatial scale that is nearly 100-fold smaller than the free space wavelength. With our chosen lattice constant of 300 nm, the normalized frequency $0.01(c/a)$ corresponds to 10 THz, i.e. a free-space wavelength of 30 μm. In other words, the guided THz photons are comparable in size to an ultraviolet photon. The graphene separation in this case is three orders of magnitude smaller than the free-space wavelength which ensures that higher order parallel-plate-waveguide modes can be neglected. If we take graphene's resistive losses into account, we observe that these extremely subwavelength modes actually have significant propagation lengths (Fig. 1(d)), and can easily reach tens of lattice constants. Across the entire Brillouin zone, the propagation length is approximately inversely proportional to the graphene scattering rate, with a more significant reduction occurring in the vicinity of the band edge where mode's group velocity is reduced.

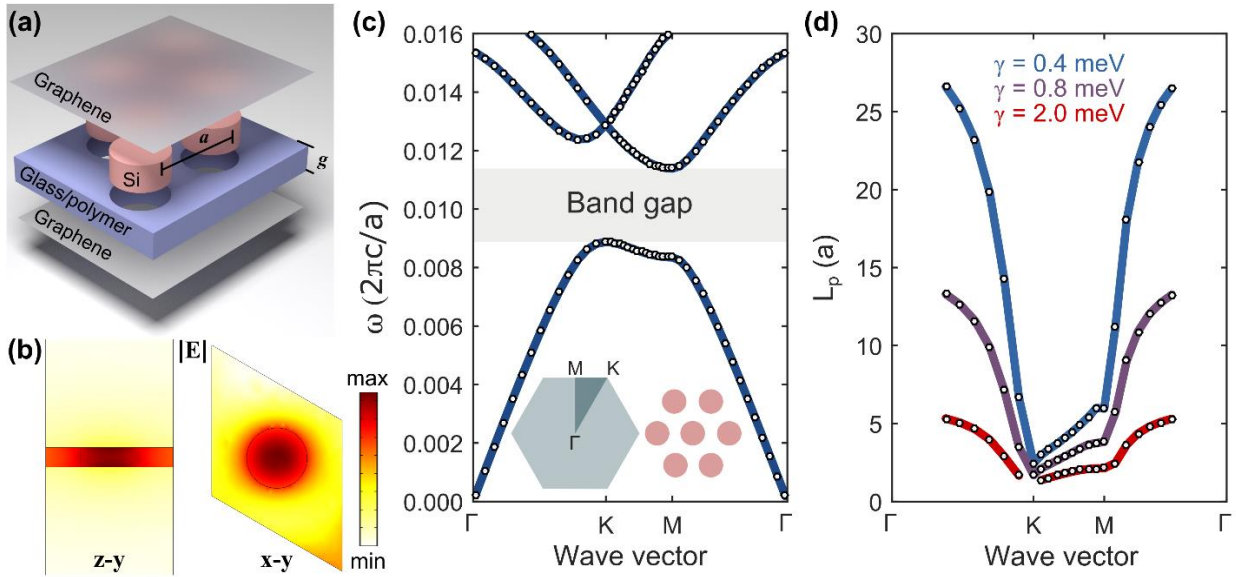

**Figure 1: Deep-subwavelength 2D photonic crystal band structures.** (a) Exploded 3D schematic of two single-layer graphene sheets sandwiching a photonic crystal slab, consisting of silicon rods $(r = 0.2a, \varepsilon_r = 11.67)$ in a triangular lattice embedded within a low-index material $(\varepsilon_r = 2.25)$. (b) Electric field magnitude of a mode at a normalized frequency of $0.007(c/a)$ on two mirror-symmetry planes in a unit cell. (c) Calculated 2D band structure along the irreducible Brillion zone boundary, with a TEM band gap (shaded in grey) from approximately $0.009(c/a)$ to $0.011(c/a)$. The left inset depicts the 2D Brillouin zone of the triangular crystal lattice. (d) Propagation length of the 1st bulk band for several graphene scattering rates. Throughout this paper, we use the following parameter values (unless otherwise noted): $E_F$ = 0.4 eV, $\gamma$ = 0.4 meV, $a$ = 300 nm, $g$ = 40 nm.

An unusual consequence of the introduction of graphene to the photonic crystal slab is that reducing the slab thickness causes a downward shift in the photonic band gap frequency range with an increase in its relative size. A large band gap is generally preferable in photonic crystal devices, because it translates to a smaller footprint from tighter field confinement, broader operational bandwidth, and stronger spatial and temporal dispersion. The thickness dependence of the gap size is shown in Fig. 2(a) where we have considered only the Γ-K and Γ-M segments of the irreducible Brillouin zone boundary because these segments define the lower and upper edges of the first band gap, respectively. The TEM band gap corresponds to the overlap between the Γ-K and the Γ-M stop gaps which is highlighted in Fig. 2(b). This scaling behavior is in stark contrast to the trend observed in index-guided photonic crystal slabs, where in slabs with thicknesses below one half the lattice constant ($g = a/2$) the modes become very weakly guided. In such index-guided structures this is observed in the band diagram with the bands shifting to higher frequencies and the band gap closing-off (if the dielectric modulation is even strong enough to support a band gap). However, in the graphene-cladded slab, the reduction in slab thickness translates to an increase in the capacitance experienced by the underlying TEM mode, which in turn, shifts the band gap to lower frequencies and enlarges its relative size as shown in Fig. 2(b) and (c). We map the location and size of the band gap as a function of thickness from approximately $g = 0.05a$ to $g = 1.25a$, where $a = 300$ nm. The minimum slab thickness considered in this range is approximately 13 nm. Fig. 2(c) shows that the relative size of the gap approaches an upper limit corresponding to the size of the band gap in the purely 2D system (approximately 28%). This relatively large band gap is made possible only by the introduction of graphene. In this frequency range, an identical photonic crystal slab *without the graphene cladding* will only support weakly guided modes with fields that extend above and below the slab and no photonic band gap (see Fig. S1).

The unique scaling behavior of the graphene-cladded slab can be explained by considering the distributed circuit model for the TEM mode, which is valid in the limit that the slab thickness is small compared with the in-plane lattice constant. The parallel capacitance between the graphene sheets is inversely proportional to $g$, whereas the kinetic inductance is independent of $g$. Since the transmission line circuit model predicts that the effective index of the bulk bands is proportional to $\sqrt{LC}$, the frequency of the band structure scales as $g^{1/2}$. We confirm that this is the case with the two dotted curves below $g/a \approx 0.125$ in Fig. 2(b), which each correspond to fits to functions of the form, $f(g) = C_0\sqrt{g}$. In the regime where the thickness is on the same order of magnitude as the lattice constant, the underlying mode becomes only quasi-TEM, and can have substantial in-plane electric field components. As the thickness continues to increase, the mode will become less uniform between the sheets and will tend to concentrate in the vicinity of the graphene sheets. At large thicknesses, the mode will no longer "feel" the intermediate photonic crystal which results in a reduced band gap. This is precisely what we observe in Fig. 2(b) and (c) above $g/a = 1.00$. In general, this thickness tuning has no precedent in all-dielectric photonic crystal slabs and may prove to be useful for the miniaturization of THz photonic circuits as it eliminates a fundamental field-confinement penalty on ultra-thin planar devices.

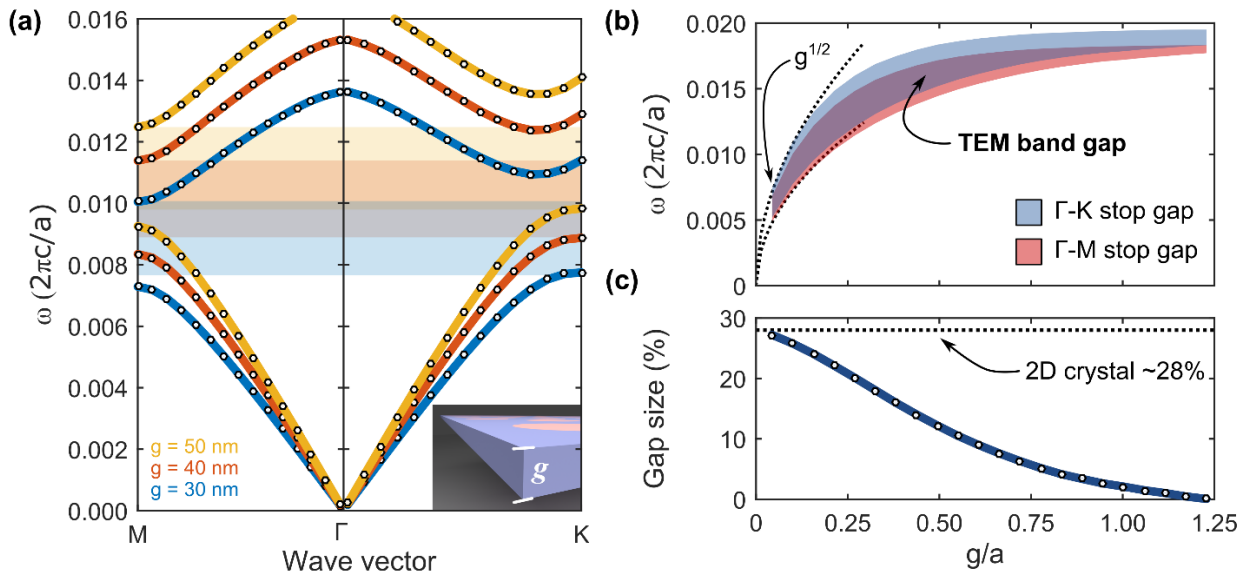

**Figure 2: Photonic band gap frequency range and size strongly dependent on the slab thickness.** (a) Bulk bands in Γ-M and Γ-K directions for several thicknesses with the TEM band gaps shaded. (b) The Γ-K stop gap, Γ-M stop gap, and TEM band gap frequency range are shaded as a function of the slab thickness. (c) Relative size of the band gap approaches that of the purely 2D crystal at small slab thicknesses and decreases significantly for larger thicknesses.

**Scaling the band structure via varying Fermi level**

Since the optical conductivity of graphene is electrostatically tunable[51,52], the band structures of the graphene-cladded photonic crystal slab can be proportionally scaled by varying the graphene Fermi level. Here we consider the graphene-cladded slab with a thickness of 40 nm and plot the first two bands under three Fermi levels: 0.25, 0.40, and 0.55 eV, in Fig. 3(a) where, again, we have only considered the Γ-K and Γ-M directions. Larger Fermi levels proportionally shift the overall band structure, including the band gap, to higher frequencies as shown in Fig. 3(b). Remarkably, the relative size of the band gap remains unchanged, as shown in Fig. 3(c), at approximately 25% for the entire range of Fermi levels considered here. This scaling behavior is consistent with the kinetic inductance's dependence on the Fermi level through the imaginary part of the surface conductivity, $\text{Im}(\sigma^{-1}/\omega) = \pi\hbar^2/e^2 E_F$.

Experimentally, graphene's Fermi level has been demonstrated[50,53] to be continuously tunable from -1 eV to 1 eV via electrostatic gating, although the Drude model given in Eq. 1 becomes inaccurate with very small absolute values of the Fermi level[53], i.e. $|E_F| < 0.05\text{eV}$. In the context of wavelength reduction, lowering the Fermi levels may appear to be desirable, because of the resultant decrease in band frequencies. However, smaller values of $E_F$ are in fact associated with larger ohmic losses. Therefore, attenuation-limited applications will impose a lower limit on the Fermi level.

Scaling the photonic band structure with the Fermi level represents a means for dynamically tuning the system after fabrication. Even though graphene-cladded photonic crystal slabs can be designed in a wide frequency range by proper choices of lattice constant and slab thickness, these geometric parameters are fixed upon fabrication. On the other hand, electrostatic gating can be performed and tuned after fabrication, using either an electrostatic field applied via external gating structures[42,43] or a bias applied across the two parallel graphene sheets[54].

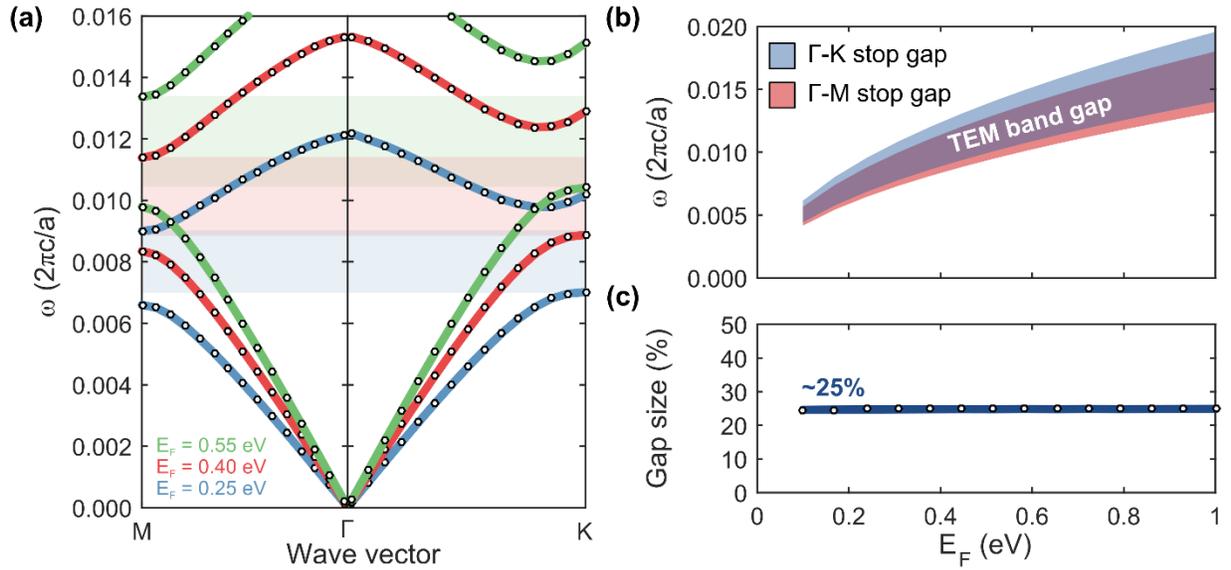

**Figure 3: The Fermi level of graphene uniformly scales the band structure and band gap spectral range.** (a) Bulk bands in $\Gamma$-M and $\Gamma$-K directions for several Fermi levels with the TEM band gaps shaded. (b) The $\Gamma$-K stop gap, $\Gamma$-M stop gap, and TEM band gap frequency range are shaded as a function of the graphene Fermi level. (c) Relative size of the band gap as a function of the graphene Fermi level.

## Scale invariance

An important consideration is whether the graphene-cladded photonic crystal slab exhibits scale invariance in the presence of graphene's dispersive Drude conductivity, which is quite different from the constant permittivity assumed in dielectric photonic crystals. At first glance, one may expect this to render our use of normalized frequencies and wave vectors invalid, however we will show that this is not necessarily the case. In the first row of Fig. 4, we plot the $\Gamma$-K stop gap, the $\Gamma$-M stop gap, and the TEM band gap of the graphene-cladded slab structure as a function of the in-plane lattice constant, $a$. The three columns correspond to a different slab thickness: $g = 40$ nm, 80 nm, and 120 nm. The second row of Fig. 4 plots the relative size of the TEM band gap for the corresponding structure in the first row, also as a function of the in-plane lattice constant.

Among the three slab thicknesses and across the range of in-plane lattice constants, we see the emergence of two distinct regimes. The first regime corresponds to the lattice constant being on *the same order of magnitude* as the thickness (the left-most portion of each column), while the second regime corresponds to the lattice constant being *much larger* than the thickness (the right-most portion of each column). Indeed, this is an expanded picture of our earlier observation and discussion of the band gap's dependence on slab thickness; now we're also considering variation of the in-plane lattice constant. To better understand the differences between these two regimes, we plot the electric field magnitude over a plane through the 3D unit cell of two specimens in the bottom row and middle column of Fig. 4. What we observe in the field profile confirms our earlier conclusion: when the thickness is on the order of the lattice constant, the field can become very non-uniform (left inset) and when the thickness is much smaller than the lattice constant, the field is highly uniform in the vertical direction (right inset). These results suggest that when the lattice constant is large relative to the thickness, the graphene-cladded slab will exhibit scale invariance with respect to its *in-plane lattice constant*, manifested as the horizontal band of the shaded regions in the first row of Fig. 4.

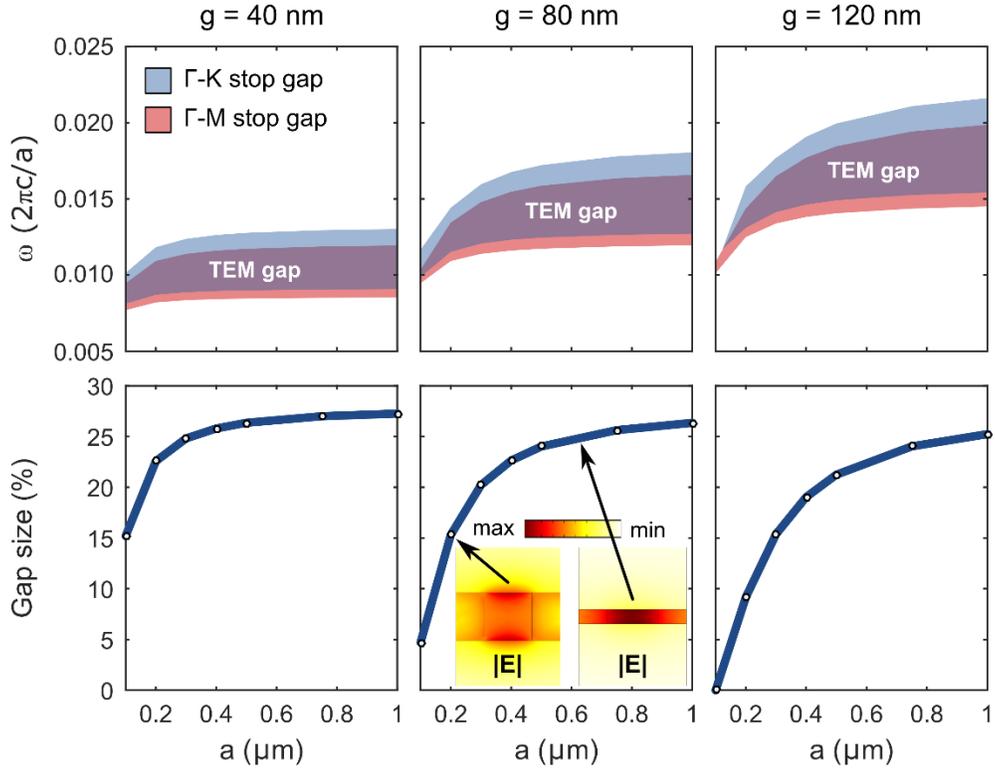

**Figure 4: Scale invariance of photonic crystal band structure.** (top row) The Γ-K stop gap, Γ-M stop gap, and TEM band gap frequency range are shaded for slab thicknesses of g = 40 nm, 80 nm, and 120 nm, as a function of the lattice constant, *a*. (bottom row) The relative size of the TEM band gap as a function of the lattice constant, for the corresponding thicknesses in the top row. Inset plots show electric field magnitude over a vertical cut (perpendicular to the direction of propagation within the crystal) through the slab with lattice constant of 200 nm and 700 nm. When the lattice constant is much larger than the thickness, the bands are scale invariant and when the lattice constant is on the order of magnitude of the thickness, the field magnitude is no longer uniform in the vertical direction (shown by the left field pattern inset), and the gap bandwidth is reduced.

Although the band gap for the TEM mode discussed so far is not a complete band gap in the strictest sense, practically it functions as a complete band gap for point and line defects, provided that the z-reflection symmetry is preserved. Only two additional modes traverse the TEM band gap (Fig. S2): the light cone which is tightly compressed to the Γ point, and the 2nd-order graphene ribbon plasmon[30], which has a field distribution with the opposite symmetry along the z direction to that of the TEM mode. In-plane imperfections, such as offsets and roughness, do not break the z-reflection symmetry and thus do not couple the TEM modes with the 2$^{nd}$-order graphene ribbon plasmons. Radiation from the TEM photonic crystal slab modes into the light cone is also suppressed for two reasons: the fields of the slab modes are tightly confined between the graphene sheets (Fig. 4), particularly at small g; the extreme subwavelength nature of the mode results in typical fabrication roughness being far smaller (1000x or more) than the free-space wavelength, causing them to be extremely inefficient dipole radiators. Thus, for practical purposes the TEM mode band gap can be considered a "complete" band gap.

**A line-defect waveguide**

Similarly to purely 2D photonic crystals, the graphene-cladded photonic crystal slab can support strongly confined guided modes in defects in the bulk crystal. For example, removing a row of rods from the crystal forms a line defect of width $\sqrt{3}a$, which can be adjusted by laterally shifting the crystal cladding on either side. The dispersion for the defect modes are presented in Fig. 5(a) for several channel widths, and closely resembles those of purely 2D photonic crystals. Another similarity to the 2D systems can be observed in Fig. 5(b), as the propagation length of the defect mode exhibits a linear dependence on its group velocity. In the widest channel considered ( $w = 1.4\sqrt{3}a$ ) both the largest group velocity and propagation length are observed. In this case the mode can propagate for more than 18 lattice constants. This propagation length is appreciably long, in the context that the guided wavelength is a factor of 40 times smaller than in free space. Much smaller propagation lengths are observed at the edges of the defect bandwidth, which is a direct result of the mode's reduced group velocity. Just like the bulk modes, the propagation length of these defect modes is ultimately limited by the graphene intrinsic scattering rate, i.e. the quality, of the graphene. In terms of the mode profile, these defect modes are strongly confined along the in-plane directions by the band gap (Fig. 5(c)) and in the out-of-plane directions by the graphene sheets. Unlike ordinary 2D photonic crystal slabs affected by the presence of radiation modes in the substrate and superstrate, the defect modes in graphene-cladded photonic crystal slabs can easily span the entire gap range with little field leakage. At very narrow channel widths we observe a flip in the sign of the group velocity.

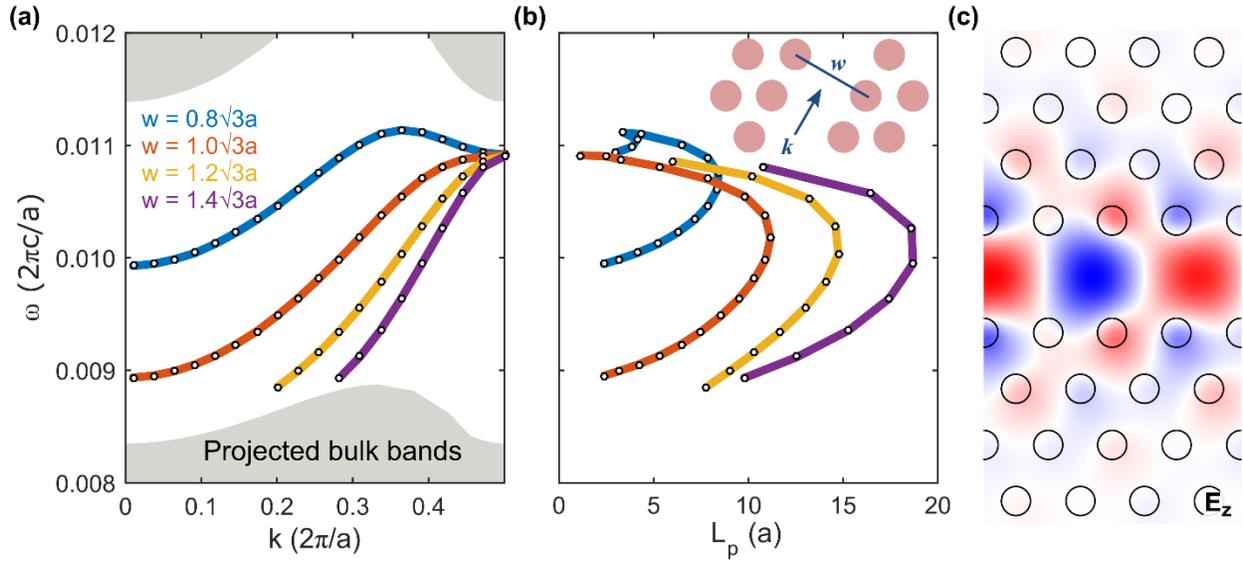

**Figure 5: Dispersion and propagation length of line-defect modes of various widths**. (a) Computed dispersion relation of line defects with a range of widths. Shaded regions are projected bands of the bulk crystal. (b) Computed propagation lengths. The inset depicts a schematic of the line defect. (c) Instantaneous out-of-plane electric field ($E_z$) of a line defect ($w = \sqrt{3}a$) at $f = 0.0100(c/a)$.

## Conclusion

In this paper we have applied graphene parallel plate waveguides to photonic crystal slabs that operate in the terahertz spectrum to achieve extremely subwavelength Bloch modes with strong out-of-plane confinement. Unlike conventional all-dielectric photonic crystal slabs, the introduction of graphene facilitates the interaction of THz radiation with feature sizes that are 100 times smaller than its free-space wavelength. This length-scale reduction is sustained in slabs that have thicknesses that are 10 or more times smaller than the in-plane crystal lattice constant. An unprecedented feature of the graphene-cladded slab is that its thickness and graphene Fermi level strongly influence the overall photonic band structure while maintaining large band gaps. The thickness, in particular, affects both the relative size and position of the gap, which is a behavior that has no precedent in all-dielectric photonic crystal slabs. The graphene Fermi level on the other hand, *uniformly* tunes the band gap position while leaving the relative bandwidth unchanged for Fermi levels between 0.1 eV and 1.0 eV. This feature allows significant flexibility in band engineering, since the band gap can be tuned via chemical doping of the graphene, or via an applied electrostatic field. Furthermore, just like in purely 2D photonic crystals, line defects in the crystal of the graphene-cladded slab can support guided modes with strong in-plane confinement and propagation lengths exceeding 10 lattice constants. Ultimately, the propagation lengths of both the bulk and defect modes of the graphene-cladded slab are limited by the ohmic losses of graphene rather than to radiation in the out-of-plane direction. In this case, the graphene quality, i.e. its intrinsic scattering rate, is the key parameter for determining the attainable transmission distances, with lower scattering rates translating to proportionally larger propagation lengths.

## Methods

All numerical results were calculated using three-dimensional finite element simulations performed in COMSOL. Eigen-frequencies and the field distributions of eigenmodes were taken from driven simulations with swept frequency and in-plane wavevector that maximize the volume-average electric energy between the graphene under constant excitation. These results were verified with a weak-form eigensolver[55] which solves for the complex-valued wavevector. The propagation length is computed from the inverse of the imaginary part of the wavevector.

## Author contributions

I.A.D.W. performed analytical and numerical modeling. S.H.M. developed the computational framework for the dispersive graphene model. Z.W. conceived and led the research. All authors contributed to data analysis and the manuscript writing.

## Acknowledgments

This work is in part supported by the Packard Fellowships for Science and Engineering, the Alfred P. Sloan Research Fellowship.